\documentclass[letterpaper,preprint,aps,floatfix,pra]{revtex4-1}
\usepackage{amsmath}
\usepackage{amssymb}
\usepackage{hyperref}
\usepackage[english]{babel}
\usepackage{graphicx}

\newcommand{\um}{-}
\begin{document}

\title{Stabilizing Canonical-Ensemble Calculations in the Auxiliary-Field
Monte Carlo Method}

\author{C. N. Gilbreth}
\email{cngilbreth@gmail.com}
\author{Y. Alhassid}
\email{yoram.alhassid@yale.edu}

\affiliation{Center for Theoretical Physics, Sloane Physics Laboratory, Yale
  University, New Haven, CT 06520, USA}

\begin{abstract}
  Quantum Monte Carlo methods are powerful techniques for studying
  strongly interacting Fermi systems. However, implementing these methods on
  computers with finite-precision arithmetic requires careful attention to
  numerical stability. In the auxiliary-field Monte Carlo (AFMC) method,
  low-temperature or large-model-space calculations require numerically
  stabilized matrix multiplication. When adapting methods used in the
  grand-canonical ensemble to the canonical ensemble of fixed particle number,
  the numerical stabilization increases the number of required floating-point
  operations for computing observables by a factor of the size of the
  single-particle model space, and thus can greatly limit the systems that can
  be studied. We describe an improved method for stabilizing canonical-ensemble
  calculations in AFMC that exhibits better scaling, and present numerical tests
  that demonstrate the accuracy and improved performance of the method.
\end{abstract}

\maketitle

\section{Introduction}

The auxiliary-field quantum Monte Carlo (AFMC) method is a widely used approach
for calculating ground-state and finite-temperature properties of interacting
quantum many-fermion systems. These include nuclei~\cite{Alhassid2012_jpcs},
condensed matter systems~\cite{Alhassid2012,LohJr1992}, atoms and
molecules~\cite{Charutz1994_jcp,Jacobi2004}, quark
matter~\cite{Ohnishi2012_lft}, and cold atomic Fermi
gases~\cite{Bulgac2008_pra,Zinner2009_pra,Carlson2011_pra,Gilbreth2013_pra}. It is one of the
few general numerical methods for such systems that takes into account
  all two-body correlations of the particles, and is exact up to a
statistical error introduced by the Monte Carlo sampling. In general, fermionic
systems can suffer from a so-called sign problem that leads to large statistical
errors.  However, for a significant class of systems, namely those with ``good sign''
  interactions, the statistical errors are reasonable, and AFMC is one of the
most robust and accurate techniques available.

Implementing AFMC at low temperatures and/or with large single-particle model
spaces can be numerically challenging. In this method, the thermal propagator
$\hat{U} =e^{- \beta \hat{H}}$, where $\hat{H}$ is the Hamiltonian of the system
and $\beta =1/T$ is the inverse temperature, is expressed as a functional
integral of a non-interacting propagator $\hat{U}(\sigma )$ parameterized by
imaginary-time-dependent fields $\sigma$.  The time dependence of these
auxiliary fields requires that $\hat{U}(\sigma)$ be represented as a long chain
of matrix multiplications, with each matrix representing the propagator for a
short time slice $\Delta\beta$. At low temperatures (i.e., large $\beta$), or
for large single-particle spaces, the repeated matrix multiplications give rise
to widely varying numerical scales, in which the intermediate and smaller scales
are hidden in the differences between much larger numbers. In finite-precision
arithmetic, these physically important intermediate and smaller scales become
unrecoverable, leading to uncontrolled numerical errors.

In the grand-canonical ensemble, this problem has a well-known
solution~\cite{LohJr1992} in which the matrix representation of
$\hat{U}(\sigma)$ in the single-particle model space is decomposed into a form
that displays the scales explicitly. However, when this stabilization method is
applied in a straightforward manner to the canonical ensemble
{\cite{Alhassid2008b}}, in which exact particle-number projection is used, it
increases the computational effort required for computing observables from $O (
N_{s}^{3} )$ to $O ( N_{s}^{4} )$, thus limiting the sizes of systems which can
be studied.

Here we describe an improved method for numerically stabilized AFMC
calculations in the canonical ensemble. This method scales as $O ( N_{s}^{3} )$
and therefore makes canonical-ensemble calculations competitive with
grand-canonical calculations. In practice it allows canonical-ensemble
calculations in spaces much larger than were previously possible.
We recently applied this method to compute signatures of the
superfluid phase transition in a finite-size cold atomic trapped Fermi
gas~\cite{Gilbreth2013_pra}. Here we describe the method in more detail and
provide evidence for its numerical stability. The method is general and is
applicable to other AFMC calculations where the canonical ensemble is important,
such as finite nuclei.

This paper is organized as follows. In Sec.~\ref{SecBackground}, we briefly
review the AFMC method and discuss the calculation of observables within the
canonical ensemble. In Sec.~\ref{SecMatmul}, we present the problem of
stabilizing matrix multiplication within AFMC and describe its standard
solution. In Sec.~\ref{SecFourier}, we discuss the calculation of observables
from the stabilized propagator, the principal subject of this work. We first
review the usual method used for the grand-canonical ensemble and then discuss
our new method for the canonical ensemble. In Sec.~\ref{secStability}, we
discuss the numerical stability and accuracy of the new method. Finally, in
Sec.~\ref{secConclusion} we present our conclusion.

\section{Auxiliary-field Monte Carlo (AFMC) method}\label{SecBackground}

The AFMC method employs a Hubbard-Stratonovich (HS)
transformation~\cite{Hubbard1959,Stratonovich1957} to rewrite the thermal
propagator $e^{- \beta \hat{H}}$ of a many-particle system as a functional
integral of a non-interacting propagator $\hat{U} ( \sigma )$ for particles
moving in imaginary-time-dependent fields $\sigma$. Here $\hat{H}$ is the
Hamiltonian, $\beta=1/T$ is the inverse temperature, and the auxiliary fields
$\sigma=\sigma (\tau)$ are functions of the imaginary time $\tau$ ($0\leq \tau
\leq \beta$).  Explicitly,
\begin{equation}
  \label{pathint} e^{- \beta \hat{H}} = \int D [ \sigma ] G ( \sigma )
  \hat{U} ( \sigma ) ,
\end{equation}
where $D [ \sigma ]$ is the integration measure, $G ( \sigma )$ is a Gaussian
weight, and $\hat{U} ( \sigma )$ is the thermal propagator for a
non-interacting system parameterized by the auxiliary fields~$\sigma(\tau)$.

The Hamiltonian $\hat{H}$ on the left-hand-side of Eq.~\eqref{pathint} is
defined in a many-particle fermionic Fock space.  This is typically generated
from a finite basis of $N_{s}$ single-particle orbitals by constructing all
possible Slater determinants in which a subset of these orbitals are
occupied. The resulting space has very large dimension, e.g., $\binom{N_{s}}{N}$
for $N$ fermions of a single species.  On the other hand, the propagator
$\hat{U}(\sigma)$ in the integrand of Eq.~\eqref{pathint} describes a
non-interacting system, and its properties can be determined by matrix algebra
in the space of single-particle states, which has much lower dimension $N_{s}$.

Observables in the AFMC method are computed by sampling their thermal
expectation values in this noninteracting system at different values of the
external fields $\sigma$. For an observable $\hat{O}$, the HS transformation
implies
\begin{equation}
  \label{traceratio} \langle \hat O \rangle = \frac{\text{Tr} (
  \hat{O} e^{- \beta \hat{H}} )}{\text{Tr} ( e^{- \beta \hat{H}}
  )} = \frac{\int D [ \sigma ] G ( \sigma ) \text{Tr} [\hat{O}
  \hat{U} ( \sigma )]}{\int D [ \sigma ] G ( \sigma ) \text{Tr} [\hat{U}(\sigma )]} \, .
\end{equation}
The traces in Eq.~\eqref{traceratio} can be computed in various subspaces of
Fock space, including the grand-canonical ensemble (where a suitable chemical
potential must also be included), and the canonical ensemble. Traces at given
values of good quantum numbers, such as spin and parity, can also be calculated
by using suitable projection
operators~\cite{Alhassid2007_prl,Nakada2008_prc,Nakada1997_prl}.

In the grand-canonical (GC) ensemble, the many-particle traces can be easily
computed from the matrix representation $U (\sigma )$ of the propagator $\hat{U}
(\sigma )$ in the single-particle space. The corresponding partition function is
\begin{equation}
  \label{zeta} \zeta ( \sigma ) = \text{Tr}_{\text{GC}} [ \hat{U} ( \sigma ) ]
  = \det [ 1+U ( \sigma ) ] ,
\end{equation}
while the expectation of a one-body operator $a_{i}^{\dagger} a_{j}$ is
\begin{equation}
  \label{aiaj} \langle a_{i}^{\dagger} a_{j} \rangle_{\sigma} =
  \frac{\text{Tr}_{\text{GC}} [ a_{i}^{\dagger} a_{j} \hat{U} ( \sigma )
  ]}{\text{Tr}_{\text{GC}} [ \hat{U} ( \sigma ) ]} = \left[ \frac{1}{1+U (
  \sigma )^{-1}} \right]_{j,i} .
\end{equation}
To compute these quantities in an $N$-particle canonical ensemble, one can apply
a discrete Fourier sum to grand-canonical
quantities~\cite{Ormand1994_prc}. This yields\footnote{To make the Fourier sum
  numerically stable a real chemical potential should be included, which
  we have omitted here for simplicity. For details, see Ref.~\cite{Ormand1994_prc}.}
\begin{equation}
  \label{zetan} \zeta_{N} ( \sigma ) = \frac{1}{N_{s}} \sum_{m=1}^{N_{s}}
  e^{-i \varphi_{m} N} \det [ 1+  e^{i \varphi_{m}} U ( \sigma ) ] ,
\end{equation}
and
\begin{equation}
  \label{aiajn} \langle a_{i}^{\dagger} a_{j} \rangle_{N, \sigma} =
  \frac{1}{\zeta_{N} ( \sigma ) N_{s}} \sum_{m=1}^{N_{s}} e^{-i \varphi_{m} N}
  \left[ \frac{1}{1+  e^{-i \varphi_{m}} U ( \sigma )^{-1}} \right]_{ji} \det [
  1+  e^{i \varphi_{m}} U ( \sigma ) ] ,
\end{equation}
where $\varphi_{m} \equiv 2 \pi m/N_{s}$. As we will see in the next section,
the presence of the Fourier sum increases the number of floating-point
operations required for numerically stabilized calculations from $O (
N_{s}^{3} )$ to $O ( N_{s}^{4} )$ when the standard method from the
grand-canonical ensemble is used.

For more details of the AFMC method and its practical applications, see
Refs.~{\cite{LohJr1992,Lang1993_prc,Alhassid1994_prl,Koonin1997_physrep,Alhassid2001_ijmpb,Bulgac2008_pra,
    Alhassid2012_jpcs,Nakada1997_prl,Alhassid2007_prl,Alhassid2008b,Nakada2008_prc,
    Gilbreth2013_pra}}.

\section{Numerical Stabilization}\label{SecMatmul}

In numerical AFMC calculations, the interval $[0,\beta]$ is divided into $N_t$
intervals of equal length $\Delta \beta =\beta/N_t$.  The $N_s\times N_s$ matrix
$U\equiv U(\sigma )$ in Eqs.~(\ref{zeta}-\ref{aiajn}) is then a time-ordered
product
\begin{equation}
  \label{uprod} U  =U_{N_{\tau}} \cdots U_{1}
\end{equation}
of $N_{t}$ factors, where $U_{k}\equiv U ( \sigma ( \tau_{k}))$ is the matrix
representation in the single-particle space of the many-particle propagator
$\hat{U}(\sigma ( \tau_{k}))$ for the $k$-th time slice. Each matrix $U_{k}$ has
the form $U_{k} =e^{- \Delta \beta h_{k}}$, where $h_{k}$ is a complex,
generally non-hermitian matrix generated from the stochastically selected fields
$\sigma (\tau_{k} )$. In AFMC calculations, the product \eqref{uprod} is
explicitly computed to obtain $U$.

At low temperature (i.e., large $\beta$), the number of factors in the product
(\ref{uprod}) becomes large. As the number of factors grows, the range of
numerical scales represented in the product $U$ diverges, and the matrix $U$
becomes ill-conditioned, i.e., has large condition number\footnote{Here the
  matrix norm $\lVert U\rVert$ is defined as $\lVert U\rVert  = \max_{x \in \mathbb{C}^{N_s}} \lVert U
  x\rVert /\lVert x\rVert $.} $\kappa(U) \equiv \lVert U\rVert  \lVert U^{-1}\rVert$. This can make it impossible
to extract information via \eqref{zetan} and \eqref{aiajn} about states in the
interior of the single-particle spectrum, as the relevant energy scales are
represented in $U$ only implicitly as the differences of much larger
numbers~\cite{LohJr1992,Koonin1997_physrep}.

The known solution to this problem is to compute a decomposed form of $U$, such
as a singular-value decomposition (SVD) or a QR
decomposition~\cite{LohJr1992,Golub1990_mc,Bai2011_laa}. These decompositions
allow an accurate floating-point representation of $U$ by keeping the widely
varying numerical scales in a separate diagonal matrix. Such a decomposition
takes the form
\begin{equation}
  \label{UADB} U=ADB= \left(\begin{array}{ccc}
    x & x & x\\
    x & x & x\\
    x & x & x
  \end{array}\right) \left(\begin{array}{ccc}
    X &  & \\
    & \scriptstyle{X} & \\
    &  & \scriptscriptstyle{X}
  \end{array}\right) \left(\begin{array}{ccc}
    x & x & x\\
    x & x & x\\
    x & x & x
  \end{array}\right) ,
\end{equation}
where $A$ and $B$ well-conditioned matrices (i.e., have condition numbers close
to $1$), and $D$ is diagonal with positive entries that represent the divergent
scales contained in $U$. In Eq.~\eqref{UADB} the size of the symbols indicates
the respective magnitude of the matrix elements. In the SVD, we decompose $U = A D B$,
where $A$ and $B$ are both unitary and the entries of $D$ are the singular
values of $U$. To use a QR decomposition, we decompose $U = Q R$, where $Q$ is
unitary and $R$ is upper triangular. We then set $A=Q$ and separate out the
diagonal matrix $D$ from $R$ by scaling the rows of $R$ such that $R$ is either
unit upper triangular~\cite{LohJr1992} or has rows with unit
norm~\cite{Bai2011_laa}. We thus obtain $U = A D B$, where $A$ is unitary, $D$ is
diagonal with positive entries, and $B$ is upper triangular. Note that for both
the SVD and QR decompositions, if we were to multiply out explicitly the factors
on the r.h.s. of Eq.~(\ref{UADB}), all resulting elements will be of the largest
magnitude, indicating that the smaller scales cannot be recovered except as
differences of much larger numbers.

To stably compute the decomposition of $U$, one first decomposes the propagator
of the first time slice, $U_{1} =A_{1} D_{1} B_{1}$, which is well-conditioned,
then updates the decomposition as the propagator for each successive time slice
is multiplied into the product (\ref{uprod})~\cite{LohJr1992}. A careful
stability analysis for this process has been performed in
Ref.~\cite{Bai2011_laa}.

\section{Stabilizing Canonical-Ensemble Calculations}\label{SecFourier}

Once the multiplication in (\ref{uprod}) is carried out stably and $U$ is
available in a decomposed form $U=A D B$ (for a particular set of fields
$\sigma$), it is necessary compute the partition function and the one-body
densities. Canonical-ensemble calculations accomplish this using the Fourier
sums \eqref{zetan} for the partition function and \eqref{aiajn} for
one-body observables. In this section, we first describe in
Sec.~\ref{standard_method} the method adapted from the grand-canonical
calculations to canonical calculations~\cite{Alhassid2008b}, and then in
Sec.~\ref{secImprovedMethod}, we describe our improved method, which is the
main subject of this paper.

\subsection{Standard method}
\label{standard_method}

The grand-canonical partition function for a particular set of fields $\sigma$
is given by the determinant $\det(1+U)$ in Eq.~\eqref{zeta}. Similar quantities
appear in the canonical partition function \eqref{zetan} as $\det ( 1+e^{i \varphi_{m}} U
)$, which must be computed for each value of $m=1, \ldots ,N_{s}$. To compute
these determinants from the decomposition $U=A D B$, one may factorize
\begin{equation}
  \label{ABfactor} 1+e^{i \varphi_{m}} ADB=A (A^{-1} B^{-1} +e^{i
  \varphi_{m}} D) B.
\end{equation}
As discussed in Ref.~\cite{LohJr1992}, the addition of $e^{i\varphi_m}D$, which
has widely diverging scales, to $A^{-1} B^{-1}$ does not introduce significant
errors into the observables~\footnote{This depends on the inclusion of a
  chemical potential to stabilize the Fourier sum~\cite{Ormand1994_prc}, which
  we have omitted here for simplicity.}. The quantity in parentheses can then be
decomposed for each $m$ as $ A^{-1} B^{-1} +e^{i \varphi_{m}} D =A_{m} D_{m}
B_{m}$, so that
\begin{equation}
  \label{detfactor} \det (1+e^{i \varphi_{m}} ADB) = \det A \det
  A_{m}  \det D_{m} \det B_{m}  \det B \;.
\end{equation}
The matrix decomposition is an $O (N_{s}^{3} )$ operation. Thus, the computation
of the canonical partition (\ref{zetan}) with this method is an $O (N_{s}^{4})$
operation, since a matrix decomposition must be performed for each value of $m$ in
the Fourier sum.

\subsection{Improved method}\label{secImprovedMethod}

If $U$ could be diagonalized by a similarity transformation, i.e., $U=P \Lambda
P^{-1}$, where $P$ is invertible and $\Lambda_{i,j} = \delta_{i,j} \lambda_{i}$
is diagonal, then the eigenvalues $\lambda_i$ may be used to compute the
determinant in $O ( N_{s} )$ operations:
\begin{equation}
  \det (1+e^{i \varphi_{m}} P \Lambda P^{-1} ) = \det (1+e^{i \varphi_{m}}
  \Lambda ) = \prod_{k=1}^{N_{s}} (1+e^{i \varphi_{m}} \lambda_{k} ) .
\end{equation}
The calculation of the Fourier sum therefore becomes an $O (N_{s}^{2} )$
operation, requiring, however, a matrix diagonalization [$O (N_{s}^{3} )$]
beforehand. Thus such a method would overall requires $O (N_{s}^{3} )$
operations. A similar estimate applies to the calculation of the one-body
densities.

However, the decomposition $U=A D B$ cannot simply be multiplied out to
diagonalize $U$, as this would destroy information contained in all but the
largest numerical scales in $D$. Instead, we can apply a simple transformation
to {\em stably} diagonalize $U$. The equation we have to solve for an eigenvalue
$\lambda$ of $U$ is
\begin{equation}
ADBx= \lambda x.
\end{equation}
Multiplying both sides by $A^{-1}$ and defining $y=A^{-1} x$, we obtain
\begin{equation}
  \label{dba} DBAy= \lambda y,
\end{equation}
where $DBA$ is a row-stratified matrix, i.e., a well-conditioned matrix $B A$
multiplied on the left by a diagonal matrix $D$ whose entries vary widely in scale
\begin{equation}
  \label{dbax} DBA= \left(\begin{array}{ccc}
    X & X & X\\
    \scriptstyle{X} & \scriptstyle{X} & \scriptstyle{X}\\
    \scriptscriptstyle{X} & \scriptscriptstyle{X} & \scriptscriptstyle{X}
  \end{array}\right) \;.
\end{equation}
The matrix $D B A$, although highly ill-conditioned, can be stably diagonalized
by first balancing the matrix and then using the QR algorithm, as is done, e.g.,
in LAPACK {\cite{lapack}}. We present numerical evidence of this in the next
section. We can then stably determine the eigenvalues and eigenvectors of $U$.
The eigenvalues of $U$ are exactly those of $D B A$, while the eigenvectors
$x_{i}$ of $U$ can be obtained from the eigenvectors $y_{i}$ of $D B A$ by the
transformation $x_{i} =A y_{i}$.

The Fourier sums \eqref{zetan} and \eqref{aiajn} can easily be
expressed in terms of the eigenvalues and eigenvectors of $U$, allowing one to
stably compute observables from the decomposition $U=A D B$ using $O (
N_{s}^{3} )$ operations. For the partition function, we have
\begin{equation}
  \zeta_{N} ( \sigma ) = \frac{1}{N_{s}} \sum_{m=1}^{N_{s}} e^{-i \varphi_{m}
  N} \prod_{k=1}^{N_{s}} (1+e^{i \varphi_{m}} \lambda_{k} ) ,
\end{equation}
while for the one-body densities, we compute
\begin{equation}
  \label{aiajn1} \gamma_{k} \equiv \frac{1}{\zeta_{N} ( \sigma ) N_{s}}
  \sum_{m=1}^{N_{s}} e^{-i \varphi_{m} N} \left( \frac{1}{1+ \lambda_{k}^{-1}
  e^{-i \varphi_{m}}} \right) \prod_{k=1}^{N_{s}} (1+e^{i \varphi_{m}}
  \lambda_{k} ) ,
\end{equation}
so that (here the $j$-th column of the matrix $P$ is the eigenvector $x_j$)
\begin{equation}
  \langle a_{i}^{\dagger} a_{j} \rangle_{N, \sigma} = \sum_{k} P_{j k}
  \gamma_{k}  P_{k i}^{-1} .
\end{equation}

\begin{figure}[ht]
\includegraphics[width=3.5in]{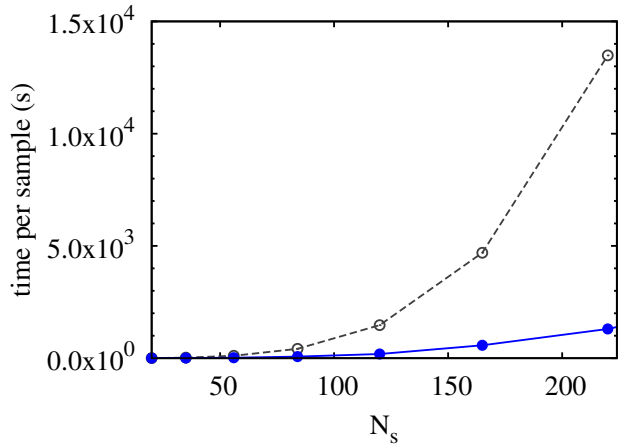}
  \caption{\label{figTiming} Comparing the timing of the standard and new
    methods for calculating observables from the stabilized propagator of
    Eq.~\eqref{UADB}. The time required to calculate one sample for a 20-atom
    cold Fermi gas at a temperature of $T=0.1 \, \hbar \omega$ ($\omega$ is the
    trap frequency) and $\Delta \beta =1/32$~\cite{Gilbreth2013_pra} is shown
    versus the number of single-particle states. Open circles describe the
    standard method of Eqs.~\eqref{ABfactor} and \eqref{detfactor}, while solid
    circles correspond to the new method using stabilized matrix
    diagonalization. Convergence for the condensate fraction in
    Ref.~\cite{Gilbreth2013_pra} was reached at $N_{\max} =11$ ($N_{s} =364$),
    which would be impractical to compute using the standard method.}
\end{figure}

We have compared the efficiency of the standard method and our improved method
in the context of a particular many-body system that is of interest to cold atom
physics. We consider two species of fermionic atoms (10 atoms of each type),
moving in an isotropic three-dimensional harmonic trap and interacting with a contact
interaction of zero range and infinite scattering length (known as the unitary
limit)~\cite{Gilbreth2013_pra}. The single-particle basis (for a given species)
consists of all eigenfunctions of the three-dimensional harmonic oscillator with
at most $N_{\max}$ oscillator quanta. The number of single-particle states for
this basis is given by $N_{s} = ( N_{\max} +1 ) ( N_{\max} +2 ) ( N_{\max} +3 )
/6$. In Fig.~\ref{figTiming} we show the time to compute one sample in AFMC
versus $N_{s}$ (the dimension of the matrices) for the standard method (open
circles) and our improved method (solid circles). We observe that the new method
yields a dramatic improvement in efficiency over the standard method. Many of
our calculations in Ref.~\cite{Gilbreth2013_pra} were done for $N_{\max} =11$
oscillator quanta, for which the number of single-particle states is $N_{s}
=364$. In the standard method, these calculations would have been too
time-consuming and thus impractical to carry out on current computers.

\section{Stability and accuracy}\label{secStability}

It is not obvious that the computation of the eigenvalues of the row-stratified
matrix $D B A$ in Eq.~\eqref{dba} is numerically stable. In fact, the standard
error analysis of the QR algorithm obtains a bound for the backward error which
is proportional to the Frobenius norm of the matrix~\cite{Tisseur1996_tr}. For
our problem, this bound is many orders of magnitude larger than the relevant
eigenvalues. This backward error also ignores the row-stratified structure of
the original matrix.

The eigenvector problem of stratified (also known as graded) matrices has been
explored previously; see, in particular, Ref.~\cite{Stewart2000_tr}. However, to
our knowledge, no proof of the stability of the QR algorithm for stratified
matrices is currently known.

This section has two parts. In Sec.~\ref{evalsens}, we first study the
sensitivity of the eigenvalues and eigenvectors of the row-stratified matrix $D
B A$ to small relative perturbations in its matrix elements. We find, using a
modification of standard perturbation theory, that the problem is
well-conditioned under essentially the same circumstances as for a matrix with a
condition number close to $1$.

In Sec.~\ref{test}, we test numerically the diagonalization of row-stratified
matrices using LAPACK. We find, when the matrix is balanced beforehand, that the
method is perfectly stable for matrices of the type considered here. We
also demonstrate the accuracy of AFMC calculations with the
improved method.

\subsection{Eigenvalue and eigenvector sensitivity}
\label{evalsens}

Let $M$ be a row-stratified matrix
\[ M=DC= \left(\begin{array}{ccc}
     X & X & X\\
     \scriptstyle{X} & \scriptstyle{X} & \scriptstyle{X}\\
     \scriptscriptstyle{X} & \scriptscriptstyle{X} & \scriptscriptstyle{X}
   \end{array}\right) , \]
where $C$ is invertible and of unit scale (i.e., has eigenvalues of order 1) and $D=
\text{diag} \{ d_{1} , \ldots ,d_{n} \}$, where $d_{1} \geqslant d_{2} \geqslant
\cdots \geqslant d_{n} >0$. (Here $M$ represents the matrix $D B A$
 and $C$ represents the product $B A$ of Sec.~\ref{secImprovedMethod}.) Suppose also
that $\lambda$ is a simple eigenvalue of $M$, and that $y$ and $z$ satisfy $M y=
\lambda y$ and $z^{\dagger} M= \lambda z^{\dagger}$.

Standard perturbation theory~\cite{Golub1990_mc} is concerned with the absolute
sensitivity of $\lambda$ with respect to small perturbations in the entries of
$M$. In particular, let $\delta M = \varepsilon E$ be a small perturbation of
$M$ representing roundoff error in the entries of $M$. Here $\varepsilon$ is on
the order of the machine precision and we may take $\lVert E\rVert _2 = \lVert M\rVert_2$ (where
$\lVert M\rVert_2 = \max_{x \in \mathbb{C}^{N_s}} \lVert M x\rVert_2/\lVert x\rVert_2$ is the matrix
2-norm). Then the standard sensitivity analysis~\cite{Golub1990_mc} gives
\begin{equation}
|\dot{\lambda}(0)| \leqslant \frac{\lVert M\rVert_2}{s(\lambda)} \,,
\end{equation}
where $\dot{\lambda}(0) \equiv (d\lambda/d\varepsilon)|_{\varepsilon=0}$ and
$s(\lambda)\equiv |z^\dagger y|$ is the condition of the eigenvalue $\lambda$.
However, this result is not particularly useful for row-stratified
matrices, since for the smaller eigenvalues, $\lVert M\rVert_2$ may be much larger than
$\lambda$.

Fortunately, we can determine a bound for the \emph{relative} error on $\lambda$
that exploits the structure of $M$. In particular, we find (see Appendix
\ref{secAppendix1})
\begin{equation}
\label{evalerr}
|\dot{\lambda}(0)/\lambda| \leqslant \frac{\lVert C^{-1}\rVert_2}{s(\lambda)},
\end{equation}
which shows that the relative sensitivity of $\lambda$ does not depend on the
condition number of $M$, but only on the condition of the matrix $C$ (which is
of unit scale) and on the condition $s(\lambda )$ of the eigenvalue $\lambda$.
We also note that, in the calculations described in this paper, $M$ is the
product of matrix exponentials and therefore all of its eigenvalues are strictly
nonzero, assuming a sufficient range in the floating-point representation.

In the SVD, $C$ is unitary, so $\lVert C^{-1}\rVert_2$ in Eq.~\eqref{evalerr} becomes 1, and
$| \dot{\lambda} (0)/ \lambda | \leqslant 1/s(\lambda)$.  On the other hand, in
the QR decomposition, $M = D R Q$, where $Q$ is unitary and $R$ is upper
triangular. In this case $C=R Q$ and Eq.~\eqref{evalerr} becomes $ |
\dot{\lambda} (0)/ \lambda | \leqslant \lVert R^{-1} \rVert_{2}/s (\lambda)$. In
practice, the matrix $R$ is well-conditioned.

Thus, we conclude that the nonzero eigenvalues of the highly ill-conditioned
but row-stratified matrix $M=D C$ are insensitive to roundoff error in $M$ when
$s (\lambda)$ is not too small. This is similar to the situation for matrices
with condition number close to 1, except that we have replaced the traditional
analysis of the absolute error~\cite{Golub1990_mc} with an analysis of the
relative error.

A similar result can also be obtained for the eigenvector sensitivity. We find
\begin{equation}
\label{evsens}
  \lVert  \dot{y}_{k} (0)\rVert_{2} \leqslant  \lVert C^{-1} \rVert_{2} \sum_{i=1,i \neq k}^{N}
  \frac{| \lambda_{k} |}{| \lambda_{k} - \lambda_{i} |} \frac{1}{s ( \lambda_{i}
    )} \,,
\end{equation}
where $y_k$ is the $k$-th eigenvector and $\dot{y}_k(0) \equiv (d y_k/d
\varepsilon) |_{\varepsilon=0}$. Thus, the sensitivity of the $k$-th eigenvector
depends on the fractional separation $| \lambda_{k} - \lambda_{i} | / |
\lambda_{k} |$ of each eigenvalue from the target eigenvalue, as well as on the
condition $s ( \lambda_{i} )$ of each eigenvalue. Again, this result is similar
to the usual result~\cite{Golub1990_mc} a for well-conditioned matrix $M$,
except that the absolute difference $| \lambda_{k} - \lambda_{i} |$ is replaced
with the relative difference $| \lambda_{k} - \lambda_{i} | / | \lambda_{k} |$
multiplied by $\lVert C^{-1} \rVert_{2}$. We see that the eigenvector sensitivity depends
only on the condition of the base matrix $C$ and of the individual eigenvalues,
and not on the condition of $M$ itself.

\subsection{Numerical verification}
\label{test}

To test the stability of diagonalizing $D B A$, we computed the eigenvalues of
a $N_{s} \times N_{s}$ complex matrix $U$ generated from a product of $N_t$ matrix
exponentials
\begin{equation}
  \label{uprod1} U=e^{- \Delta \beta  h_{}} \cdots e^{- \Delta \beta  h}
  \hspace{1em} \left( N_{t}   \text{  \text{times}} \right) ,
\end{equation}
where $h$ is a randomly generated matrix (identical in each factor).  The matrix
$U$ is ill-conditioned and its calculation requires stabilized matrix
multiplication. However, its eigenvalues and eigenvectors can be determined
accurately from a single factor $e^{- \Delta \beta h}$, which is
well-conditioned. Hence, such a matrix provides a convenient test for the
diagonalization of matrices of the type that occur in AFMC.

For this test we chose the entries $h_{i j}$ to be complex numbers whose real
and imaginary parts are randomly drawn from a uniform distribution on $( 0,1
)$. This type of matrix simulates the kind that occurs in the AFMC method. We
computed $U$ in two different ways: (i) using unstabilized matrix multiplication
(with the BLAS routine ZGEMM~\cite{Dongarra1990_toms}), and (ii) using matrix
multiplication stabilized with a QR decomposition.  We then compared the
eigenvalues and eigenvectors obtained by diagonalizing $U$ with those obtained
by diagonalizing a single factor $e^{- \Delta \beta h}$. In exact arithmetic,
the eigenvectors of $U$ should be identical to those of $e^{- \Delta \beta h}$,
while the eigenvalues should be the $N_{t}$-th power of those of $e^{- \Delta
  \beta h}$.

\begin{figure}[ht]
  \includegraphics[width=3.5in]{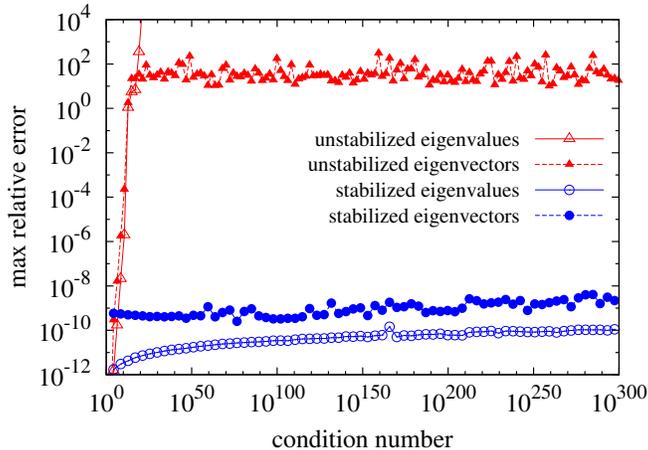}
  \caption{\label{figerr}Relative errors in the eigenvalues and eigenvectors of
    a row-stratified, ill-conditioned matrix $U$ generated from a long product
    of matrix exponentials \eqref{uprod1}. The horizontal axis shows the
    approximate condition number of $U$ (computed as the magnitude of the ratio
    of the largest to smallest eigenvalues). On the vertical axis are the
    relative errors $| \lambda - \lambda_{\text{exact}} | / |
    \lambda_{\text{exact}} |$ and $\lVert  v-v_{\text{exact}} \rVert / \lVert 
    v_{\text{exact}} \rVert$ of the eigenvalues and eigenvectors of $U$,
    respectively. Open triangles: eigenvalue errors from unstabilized matrix
    multiplication (using ZGEMM) and diagonalization; solid triangles:
    eigenvector errors from the same method. Solid circles: eigenvector errors
    from QR-stabilized matrix multiplication and diagonalization of $D R Q$;
    open circles: eigenvalue errors from the same method. We observe that for
    these matrices the QR stabilization method is perfectly stable up to
    condition numbers of $10^{300}$.}
\end{figure}

We show in Fig.~\ref{figerr} the numerical relative error in the eigenvalues and
eigenvectors of $U$ as a function of the condition number~\footnote{In this test
  the condition number is estimated from
  $|\lambda_{\text{max}}/\lambda_{\text{min}}|$, i..e, the absolute value of the
  ratio of the largest-magnitude eigenvalue to the smallest-magnitude
  eigenvalue.} of $U$, which grows monotonically with the number $N_{t}$ of
factors. As the figure shows, the QR stabilization method together with
diagonalization of $D B A$ is perfectly stable for products of the form
(\ref{uprod1}).

To illustrate the accuracy of the AFMC calculations performed with the new
stabilization method, we used again the example of the cold atom Fermi gas with
contact interaction in a harmonic trap (for 20 particles). In
Table~\ref{tabNumerics} we show a comparison of the expectation values $\langle
\hat{H} \rangle$ of the Hamiltonian using the standard and new stabilization
methods, averaged over two samples.  We also list the times required to compute
these samples. The temperature used here ($T=0.1 \hbar \omega$) is well within
the region where stabilization is necessary. The results in
Table~\ref{tabNumerics} demonstrate that the two methods are numerically
identical to a large number of digits.

\begin{table}[ht]
   \begin{tabular}{cccccc}
    \hline
    & & \multicolumn{2}{c}{Standard method} & \multicolumn{2}{c}{New method} \\
    \hline
    $N_{\max}$ & $N_{s}$ & $\langle \hat{H} \rangle$ $( \hbar
    \omega )$ &  time $( s )$ & $\langle \hat{H}
    \rangle$ $( \hbar \omega )$ & time $( s )$\\
    \hline
    3 & 20 & 47.14058966 & 8.6 & 47.14058966 & 2.1\\
    4 & 35 & 47.38511486 & 48.3 & 47.38511486 & 7.4\\
    5 & 56 & 44.25256873 & 212.5 & 44.25256873 & 22.0\\
    6 & 84 & 46.36694696 & 836.2 & 46.36694696 & 69.4\\
    7 & 120 & 45.31182132 & 2941.4 & 45.31182132 & 184.3\\
    8 & 165 & 45.80579528 & 9386.9 & 45.80579528 & 575.0\\
    9 & 220 & 37.97768943 & 26973.2 & 37.97768943 & 1300.4\\
    \hline
  \end{tabular}

  \caption{\label{tabNumerics} Energies and timings for two samples of a Monte
    Carlo simulation of a 20-atom, three-dimensional isotropically trapped cold
    atomic Fermi gas at $T=0.1 \hbar \omega$. The atoms interact strongly with a
    contact interaction in the unitary limit of infinite scattering length. In
    the first and second columns (from the left) we list the maximal number
    $N_{\max}$ of oscillator quanta and the corresponding number $N_{s}$ of
    single-particle states. The third and fourth columns are, respectively, the
    expectation $\langle \hat{H} \rangle$ of the Hamiltonian averaged over two
    samples, and the time required to compute these two samples using the
    standard stabilization method. The fifth and sixth columns are the same
    quantities but using the new stabilization method. The numbers for $\langle
    \hat{H} \rangle$ shown here are not physical (as they are calculated from
    only two samples), but they clearly demonstrate that the two methods give
    numerically identical results.  The times per sample for both methods are
    also shown in Fig.~\ref{figTiming}.}
\end{table}

\section{Conclusion}\label{secConclusion}

Numerically stabilized calculations of observables for a non-interacting
propagator $\hat{U}$ are critical to performing AFMC calculations at low
temperature and/or in large single-particle model spaces.  We have described an
improved method for computing the particle-number-projected partition
function and the expectation values of observables from a stabilized matrix
decomposition $U = A D B$ of the propagator. The method works by employing a
stabilized matrix diagonalization method for $U$ and computing the partition
function and observables from the eigenvalues and eigenvectors of $U$. This new
method reduces the $O(N_{s}^{4})$ scaling of the standard method (when applied
to the canonical ensemble) to $O(N_{s}^{3})$ (where $N_s$ is the number of
single-particle states).  We have demonstrated that the new method can
dramatically reduce the computational time of canonical AFMC calculations in the context
of a trapped cold atom Fermi system. The method is also applicable to other
physical systems such as nuclei, and can enable the study of systems in the
canonical ensemble which previously could not be practically studied using AFMC.

The method relies on the stable diagonalization of an ill-conditioned but
row-stratified matrix that arises as a product of matrix exponentials. We
studied the perturbation theory for this problem and found that it is
well-conditioned (i.e., insensitive to roundoff in the input matrix) under the
same circumstances as for a well-conditioned matrix. Moreover, we demonstrated
in numerical tests that the QR algorithm (as employed by LAPACK, in which it is
preceded by matrix balancing) is numerically stable for this problem. This
method may also apply to other calculations where information must be extracted
from a dense, highly stratified matrix.

\section*{Acknowledgements}

This work was supported in part by the Department of Energy grant
DE-FG-0291-ER-40608. Computational cycles were provided by the NERSC high performance computing facility at LBL, and by the facilities of the Yale University Faculty of Arts and Sciences High Performance Computing Center. The acquisition of these facilities was partially funded by the National Science Foundation under grant No.~CNS 08-21132.

\section*{Appendix A}

\label{secAppendix1}

To determine the relative sensitivity of the eigenvalues to roundoff error in
the entries of $M$, we consider a perturbation of the form $\delta M=
\varepsilon D F$, where $\lVert F\rVert_2 =1$ (here $\lVert F\rVert_2 \equiv \text{max}_{x \ne 0}
\lVert F x\rVert_2/\lVert x\rVert_2)$ and $\varepsilon$ is small (on the order of the machine
precision). Following Ref.~\cite{Golub1990_mc}, there exist differentiable
$x(\varepsilon)$ and $\lambda(\varepsilon)$ in a neighborhood of $\varepsilon
=0$ such that
\[ (M +  \varepsilon D F) x ( \varepsilon ) = \lambda ( \varepsilon ) x (
   \varepsilon ) . \]
To determine the sensitivity of $\lambda$ with respect to the perturbation, we
differentiate both sides with respect to $\varepsilon$ and set $\varepsilon
=0$. It is easy then to see that
\begin{equation}
  \label{err1} |\dot\lambda(0)| =  \frac{|y^{\dagger} D F
  x|}{|y^{\dagger} x|} = \frac{|y^{\dagger} D F x|}{s ( \lambda )} ,
\end{equation}
where $\dot\lambda(0)\equiv (d\lambda/d\varepsilon)|_{\varepsilon=0}$ and $s (
\lambda ) \equiv |y^{\dagger} x|$ is the condition of the eigenvalue
$\lambda$. It then follows that
\begin{eqnarray*}
  | \dot{\lambda} (0)| & = & \frac{|y^{\dagger} D C C^{-1} F x|}{s ( \lambda
  )}\\
  & = &  \frac{| \lambda |  | y^{\dagger} C^{-1} F x |}{s ( \lambda )} .
\end{eqnarray*}
Using the fact that $\lVert  y_{2} \rVert = \lVert x\rVert_{2} =1$ and $\lVert F\rVert_{2} =1$, we obtain
\begin{equation}
  | \dot{\lambda} (0)/ \lambda | \leqslant \frac{\lVert C^{\um
  1} \rVert_{2}}{s ( \lambda )} \,.
\end{equation}

To obtain Eq.~\eqref{evsens} we follow a similar procedure based on the analysis
of Ref.~\cite{Golub1990_mc}.

\bibliography{bibliography}

\end{document}